\documentclass[english,journal=jaccck,manuscript=article,layout=twocolumn]{achemso}
\usepackage[T1]{fontenc}
\usepackage[utf8]{inputenc}
\usepackage{color}
\usepackage{babel}
\usepackage{array}
\usepackage{textcomp}
\usepackage{multirow}
\usepackage{amsmath}
\usepackage{graphicx}
\usepackage[unicode=true,pdfusetitle,
 bookmarks=true,bookmarksnumbered=false,bookmarksopen=false,
 breaklinks=false,pdfborder={0 0 0},pdfborderstyle={},backref=false,colorlinks=true]
 {hyperref}
\hypersetup{
 allcolors=blue}

\makeatletter

\title{Sub-Band-Gap Absorption Mechanisms Involving Oxygen Vacancies in
Hydroxyapatite}

\author{Vladimir S. Bystrov}

\affiliation[Russian Academy of Sciences]{Institute of Mathematical Problems of Biology, Keldysh Institute
of Applied Mathematics, Russian Academy of Sciences, Vitkevicha street
1, Pushchino, 142290, Moscow region, Russian Federation}

\author{Leon A. Avakyan}

\affiliation[Southern Federal University]{Physics Faculty, Southern Federal University, Zorge street 5, Rostov-on-Don
344090, Russian Federation}

\author{Ekaterina V. Paramonova}

\affiliation[Russian Academy of Sciences]{Institute of Mathematical Problems of Biology, Keldysh Institute
of Applied Mathematics, Russian Academy of Sciences, Vitkevicha street
1, Pushchino, 142290, Moscow region, Russian Federation}

\author{José Coutinho}

\affiliation[University of Aveiro]{Department of Physics \& I3N, University of Aveiro, Campus Santiago,
3810-193 Aveiro, Portugal}

\email{jose.coutinho@ua.pt}

\phone{+351 234 247051}

\fax{+351 234 378197}

\providecommand{\tabularnewline}{\\}

\setkeys{acs}{articletitle=true}

\makeatother

\begin{document}
\begin{abstract}
Hydroxyapatite (HAp) is a widely used biomaterial for the preparation
of bone and dental implants. Despite the relevance of HAp in medicine,
exciting applications involving this material as a bio-compatible
photocatalyst, depend on how well we understand its fundamental properties.
Experimental evidence suggests that oxygen vacancies play a critical
role in the production of surface radicals upon exposure of HAp to
ultra-violet (UV) light. However, very little is known about the underlying
physical and chemical details. We present a hybrid density-functional
study of the structural and electronic properties of oxygen vacancies
in large HAp supercells within the plane-wave formalism. We find that
under equilibrium conditions, vacancies occur either as a simple vacant
oxygen site (in the neutral charge state), or as extended structures
replacing several crystalline moieties (in the double plus charge
state). Large atomic relaxations upon ionization make the oxygen vacancy
a negative-$U$ defect, where the single plus charge state is metastable,
being only accessible under UV excitation. From inspection of the
transition levels, we find that electron promotion from the valence
band top to the donor state of the vacancy, involves a zero-phonon
transition of 3.6-3.9~eV. This mechanism is the most likely explanation
to the 3.4-4.0~eV absorption onset for the observation of photocatalysis
using HAp under persistent UV illumination. \emph{Published
in Journal of Physical Chemistry C, \textbf{123}, 4856-4865 (2019)}.
\texttt{\textbf{\textcolor{blue}{https://doi.org/10.1021/acs.jpcc.8b11350}}}
\end{abstract}

\section{Introduction}

Hydroxyapatite (HAp) {[}Ca$_{10}$(PO$_{4}$)$_{6}$(OH)$_{2}${]}
is the main mineral component of bone tissue and teeth, along with
the organic component (collagen) and living bone cells (osteoclasts,
osteoblasts and osteocytes).\cite{Kay1964,Young1967,Elliot1994,Currey2002,Koester2008}
HAp crystallizes within the gaps of stacked tropocollagen fibrils,
forming and strengthening the bone structure.\cite{Weiner1986,Kanzaki1998,Currey2002,Koester2008}

Upon bone fracture/destruction, bone repair occurs through simultaneous
formation of the organic base and crystallization of HAp in the form
of platelets or needle-like nanocrystals.\cite{Weiner1986,Kanzaki1998,Currey2002,Koester2008,Bystrov2011}
Such innate bio-activity and compatibility, makes HAp a widely used
material in medicine, with applications in bone and dental implants,
often integrating and covering other stiffer materials in order to
meet specific mechanic requirements.\cite{Ratner2004,Best2008,Leon2009}

It is important to distinguish between mineral (stoichiometric) from
biological HAp (bio-HAp) found in living organisms. The latter differs
from the ideal crystal due to stoichiometric imbalances in the form
of defects (\emph{e.g.} oxygen vacancies, OH-group vacancies and interstitial
protons) or even impurities, among which carbonate ions are the most
relevant. The importance of defects is paramount --- besides polarizing
the surface of bio-HAp, which is critical for adhesion with living
tissues and bone regeneration,\cite{Kobayashi2001,Nakamura2009,Slepko2013,Zeglinski2014}
defects are also responsible for several physical properties, promising
exciting functionalities to HAp. For instance, due to its peculiar
OH-channels, protons were suggested to travel long distances under
the influence of electric fields, thus enabling the production of
a “persistent polarization”.\cite{Nakamura2001,Bystrov2009,Horiuchi2012}

Tofail \emph{et~al.}\cite{Tofail2005} interpreted the polarization
of HAp as arising from a collective alignment of individual OH anions.
Accordingly, an anti-ferroelectric to ferroelectric transition was
found at 210~ºC by monitoring the temperature dependence of the dielectric
constant of HAp. Particularly elucidating experiments were described
by Hitmi and co-workers\cite{Hitmi1986,Hitmi1988} regarding the flipping
dynamics of OH$^{-}$ anions in F$^{-}$ or Cl$^{-}$ heavily doped
material. Accordingly, thermal stimulated current measurements were
employed to monitor the polarization changes in the material as the
temperature was progressively increased. Interestingly, they interpreted
that the 210~ºC phase change as an entangled flipping of OH-chains,
which was hindered in doped material due to formation of $\mathrm{F}^{-}\cdots\mathrm{H}^{+}\mathrm{O}^{2-}$
or $\mathrm{Cl}^{-}\cdots\mathrm{H}^{+}\mathrm{O}^{2-}$ hydrogen
bonds.

The collective interactions between OH dipoles confer to HAp properties
like ferroelectricity, pyroelectricity and piezoelectricity,\cite{Lang2011,Lang2013,Lang2016}
making HAp a strong contender for charge storage applications\cite{Tofail2015}
as well as to fabricate bio-electronic devices for \emph{in-vivo}
and \emph{ex-vivo} applications.\cite{Tofail2016}

Another feature exhibited by HAp is the capability of degrading hazardous
organic and inorganic chemicals, both in air and water, upon exposure
to ultra-violet (UV) light.\cite{Nishikawa2002,Nishikawa2004,Ozeki2007,Reddy2007,Tsukada2011,Shariffuddin2013,Brazon2016,Piccirillo2017}
In fact, HAp was demonstrated to rival against powerful and well-established
photo-catalysts such as TiO$_{2}$. However, unlike the latter, HAp
has the advantage of showing strong affinity for organic compounds
and living organisms, bringing obvious advantages for instance if
used for environmental remediation or cancer treatment.\cite{Wakamura2003,Hu2007,Xu2011}
Again, defects and dopants play a central role here. Electron spin
resonance (ESR) studies clearly show that the surface reactive species
is superoxide O$_{2}^{\bullet-}$.\cite{Nishikawa2002,Nishikawa2007}
This radical is formed when adsorbed oxygen captures an UV-excited
electron from a defect, most probably an oxygen vacancy.\cite{Nishikawa2002,Nishikawa2007}
The photocatalytic activity of HAp seems to be thermally controllable
--- heat treatments around 200~°C resulted in efficient photo-degradation
of several pollutants, while treating the same powder at 1150~°C
led to photo-inactive material.\cite{Nishikawa2002} This effect was
correlated with the detection of O$_{2}^{\bullet-}$ and OH$^{\bullet}$
radicals by ESR.\cite{Nishikawa2003b} It is not clear why annealing
at a moderate temperature results in photo-active HAp, whereas high-temperature
treatments deactivates photocatalysis. It is known that high temperature
treatments improve crystallinity but also result in the loss of OH$^{-}$
anions. Given the above annealing temperatures, it is tempting to
relate the activation of the photocatalytic activity after 200~°C
heat treatments, with the motion/polarization of OH-anions (also taking
place at about this temperature\cite{Tofail2005}). Unfortunately,
the available data in the literature is too scarce to draw any definite
conclusion. Further experiments would be important to clarify this
aspect, for instance by simultaneously monitoring the crystalline
structure and photo-catalysis activity of samples after being subject
to different thermal treatments.

In a latter report, using X-ray photoemission (XPS) and Fourier-transform
infra-red (FTIR) spectroscopy, Nishikawa\cite{Nishikawa2007} found
that the low-temperature annealed material (200~°C) showed some changes
in the phosphate groups upon UV illumination, suggesting that a perturbation
of P-O bonds were connected to the production of O$_{2}^{\bullet-}$
surface radicals. A drastic decrease in the intensity of several absorption
bands related to P-O vibrations was actually observed after exposing
as-grown HAp samples to UV light for 1~hour.\cite{Reddy2007,Shariffuddin2013}
The IR spectrum regained its original shape after leaving a used HAp
sample in darkness for 1~hour, demonstrating that the UV-induced
transformation of the P-O bonds was reversible.\cite{Reddy2007}

An important question is therefore: what are the details of the physical
and chemical mechanisms behind these photo-catalytic processes? It
is clear that oxygen vacancies and related defects play a crucial
but yet unknown role, which will stay undisclosed while their electronic
and optical properties are largely unexplored.

Recently, several point defects in HAp were inspected by means of
local density density functional theory, where it was found that an
oxygen vacancy in the phosphate (PO$_{4}$) unit introduces a fully
occupied state in the lower half of the gap.\cite{Bystrov2015} The
photoionization of this level was actually claimed to account for
the photocatalytic threshold in the range 3.4-4.0~eV.\cite{Bystrov2016}
However, the lack of physical significance of the Kohn-Sham eigenvalues,
along with large self-interaction errors from the local density approximation,
make any assignment of optical data a highly speculative exercise.\cite{Martin2016}

In order to obtain an accurate picture of the electronic and optical
properties of defects in a material like HAp, it is crucial to accurately
describe the physics of the material itself, which includes properties
such as the crystalline structure, the electronic band structure,
dielectric response, among others.

In a recent work, we have shown how the quality of the exchange-correlation
treatment affects the above properties.\cite{Avakyan2018} We found
out that the use of hybrid density functional theory (hybrid-DFT),
which mixes a portion of Fock exchange with a (semi-)local exchange-correlation
potential, leads to electronic properties very close to those obtained
using highly accurate many-body perturbation theory within the $GW$
framework. While the $GW$ and hybrid-DFT band gaps were close to
7~eV, generalized gradient approximated results underestimated this
figure by 30\%.\cite{Calderin2003,Rulis2004,Rulis2007,Matsunaga2007,Slepko2011,Bystrov2015}
Comparable errors are therefore expected for the calculated electronic
and optical properties of defects, as shown for the case of the OH
vacancy in HAp.\cite{Avakyan2018} From the experimental perspective,
measurements of the band gap width of HAp have been controversial.
Whereas combined photoluminescence and surface photovoltage spectroscopy
studies found a band gap approaching 4~eV,\cite{Rosenman2007} diffuse
reflectance spectra in the UV--VIS region indicate that the gap should
be wider than 6~eV,\cite{Tsukada2011} thus confirming the $GW$
results.\cite{Avakyan2018} We can only explain the narrower photoluminescent
gap with the presence of defects with levels close to mid-gap. Further,
we known that (semi-)local DFT severely underestimates the band gap
width of semiconductors and insulators. A typical generalized gradient
approximated calculation gives $E_{\mathrm{g}}=5.2$~eV (see Ref.~\citenum{Avakyan2018}
and references therein), making the reported gap of 4~eV unrealistic.

In view of the above, we present below a theoretical study of the
electronic properties of oxygen vacancies in HAp, which allows us
to connect their structure stability, charge state transitions and
photoionization energies with the optical excitation needed for the
observation of efficient photocatalytic activity.

\section{Methods\label{sec:method}}

The calculations were carried out using density functional theory
(DFT), as implemented by the Vienna Ab-initio Simulation Package (VASP).\cite{Kresse1994,Kresse1996,Kresse1999}
The exchange-correlation potential was evaluated either using the
generalized gradient approximation according to Perdew, Burke and
Ernzerhof (PBE),\cite{Perdew1996} or the Becke three-parameter, Lee-Yang-Parr
(B3LYP) hybrid functional,\cite{Lee1988,Becke1993} which incorporates
a fraction of exact exchange with local and semi-local functionals.
Due to severe limitations of the (semi-)local functionals alone in
describing the electronic structure of HAp near the band gap,\cite{Avakyan2018}
the use of hybrid-DFT turns out to be critical in the calculation
of the vacancy transition levels. Core states were described by means
of the projector augmented wave (PAW) method,\cite{Bloechl1994} while
the Kohn-Sham problem was addressed by using plane-waves with kinetic
energy up to $E_{\mathrm{cut}}=400$~eV to expand the wave functions.
For further details, including convergence tests, we direct the reader
to Ref.~\citenum{Avakyan2018}.

Oxygen vacancies were introduced in hexagonal supercells made up of
$2\times2\times2=8$ polar HAp unitcells (spacegroup $P6_{3}$ and
\#173 in the crystallographic tables), comprising a total of 352 atoms.
Although the monoclinic phase $P2_{1}/b$ was found to be more stable,\cite{Haverty2005,Ma2009}
they only differ by a few tens of meV per unit cell, with their electronic
structures being essentially identical \cite{Slepko2011}. The equilibrium
lattice parameters of $P6_{3}$-symmetric HAp were $a=9.537$~Å and
$c=6.909$~Å ($a=9.577$~Å and $c=6.877$~Å) as obtained within
PBE (B3LYP) level calculations. These compare well with the experimental
figures $a=9.417$~Å and $c=6.875$~Å.\cite{Hughes1989} A full
relaxation of such large supercells using plane-wave hybrid-DFT is
prohibitively expensive. Instead, defect structures were first found
by relaxing all atomic coordinates within PBE, until the maximum force
became less than 10~meV/Å. The resulting structures were employed
on a second step, where the total energy was obtained within hybrid-DFT
by means of a single-point calculation. This procedure was necessary
due to the sheer size of the Hamiltonian at hand combined with the
use of a plane-wave method. We have recently shown that relative energies
obtained within this methodology are usually affected by error bars
of the order of 10~meV or lower.\cite{Coutinho2017,Gouveia2019}
For instance, the energy difference between HAp unit cells with $P6_{3}$
and $P6_{3}/m$ symmetry (the latter being obtained by aligning the
OH units along opposite directions) is $E(P6_{3}/m)-E(P6_{3})=0.39$~eV
with both energies being obtained from fully relaxed B3LYP-level calculations.
An analogous quantity obtained from energies calculated according
to the two-step recipe described above, differed by less than 5~meV
from this figure. We also calculated the energy difference between
oxygen vacancies of type I and IV (see beginning of Section~\ref{sec:results}).
For practical reasons, these calculations were carried out by placing
the defects in single unit cells. While fully relaxed B3LYP-level
calculations give 1.37~eV (in favor of the O(IV) vacancy), the B3LYP
single-point calculation (using the PBE-relaxed structure) gives 1.31~eV,
also favoring the O(IV) vacancy.

For the HAp unit cell, we have shown previously that convergence of
the electron density and energy is obtained when sampling the band
structure with a $\Gamma$-centered $2\times2\times3$ mesh of \textbf{k}-points
within the first Brillouin zone (BZ). By doubling the size of the
cell along all principal directions, reciprocal lattice vectors are
contracted by a factor of two, so that a $1\times1\times2$ $\mathbf{k}$-point
grid would actually improve on the sampling quality. From convergence
tests, we found that the total energy of the 352-atom bulk supercell
obtained with a $1\times1\times1$ grid ($\mathrm{\mathrm{\Gamma}}$-point
sampling) differs by less than 0.1~eV from a $1\times1\times2$-sampled
calculation. More importantly, relative energies and ionization energies
of defects differ by about 1~meV only. Therefore, all defect calculations
employed a $\Gamma$-point sampling.

The formation energy of a defect in a crystalline sample (with arbitrary
volume) can be written as,\cite{Qian1988}

\begin{equation}
E_{\mathrm{f}}=E(q,R)-E_{\mathrm{HAp}}-\sum_{i}\Delta n_{i}\mu_{i}+q\left(E_{\mathrm{v}}+E_{\mathrm{F}}\right),\label{eq:formation}
\end{equation}
which depends primarily on the energy difference between defective
and pristine crystals, namely $E$ and $E_{\mathrm{HAp}}$, respectively.
The defect is considered on a particular configuration $R$ and charge
state $q$. The third and fourth terms account for any stoichiometric
and charge imbalances between the first two terms. Accordingly, $\mu_{i}$
are chemical potentials for any species $i$ which are respectively
added ($\Delta n_{i}$ > 0) or removed ($\Delta n_{i}$ < 0) from
or to the perfect crystal the make the defect. The last term accounts
for the energy involved in the exchange of electrons between the defect
in charge state $q$ and an \emph{electron reservoir} with a chemical
potential $\mu_{e}$ = $E_{\mathrm{v}}$ + $E_{\mathrm{F}}$, where
$E_{\mathrm{v}}$ and $E_{\mathrm{F}}$ stand for the energy of the
valence band top and Fermi level, respectively. The latter mostly
depends on the type and amount of defects and impurities in the crystal,
and can vary in the range $0\leq E_{\mathrm{F}}\leq E_{\mathrm{g}}$,
where $E_{\mathrm{g}}$ is the band gap width of the material.

For an oxygen vacancy, $\Delta n_{\mathrm{O}}=-1$ and $\Delta n_{i}=0$
otherwise, so that the third term in Eq.~\ref{eq:formation} reduces
to the oxygen chemical potential ($\mu_{\mathrm{O}}$). In the present
work we are mainly interested in the electronic properties of the
oxygen vacancy, which are independent of the choice of the chemical
potentials. Hence, we simply consider $\mu_{\mathrm{O}}$ as the energy
per atom in molecular oxygen. This was calculated by placing a molecule
in a box (edge length 20~Å) in the spin-1 ground state.

Finally, we note that because we are using periodic boundary conditions,
the calculation of a defect with net charge $q$ within the supercell
would lead to a diverging total energy. This is avoided by spreading
a uniform background counter-charge with density $\rho_{\mathrm{back}}=-q/\Omega$
over the cell volume $\Omega$.\cite{Makov1995} The spurious Coulomb
interactions between the periodic array of charged defects and background
must be removed from the periodic total energy $\widetilde{E}$, so
that the energy of the defective crystal in Eq.~\ref{eq:formation}
should be replaced by $E=\widetilde{E}+E_{\mathrm{corr}}$, where
the correction term $E_{\mathrm{corr}}$ was obtained according to
the method proposed by Freysoldt, Neugebauer and Van de Walle \cite{Freysoldt2009},
and recently generalized for anisotropic materials.\cite{Kumagai2014}

\section{Results\label{sec:results}}

\begin{figure*}
\includegraphics[width=17cm]{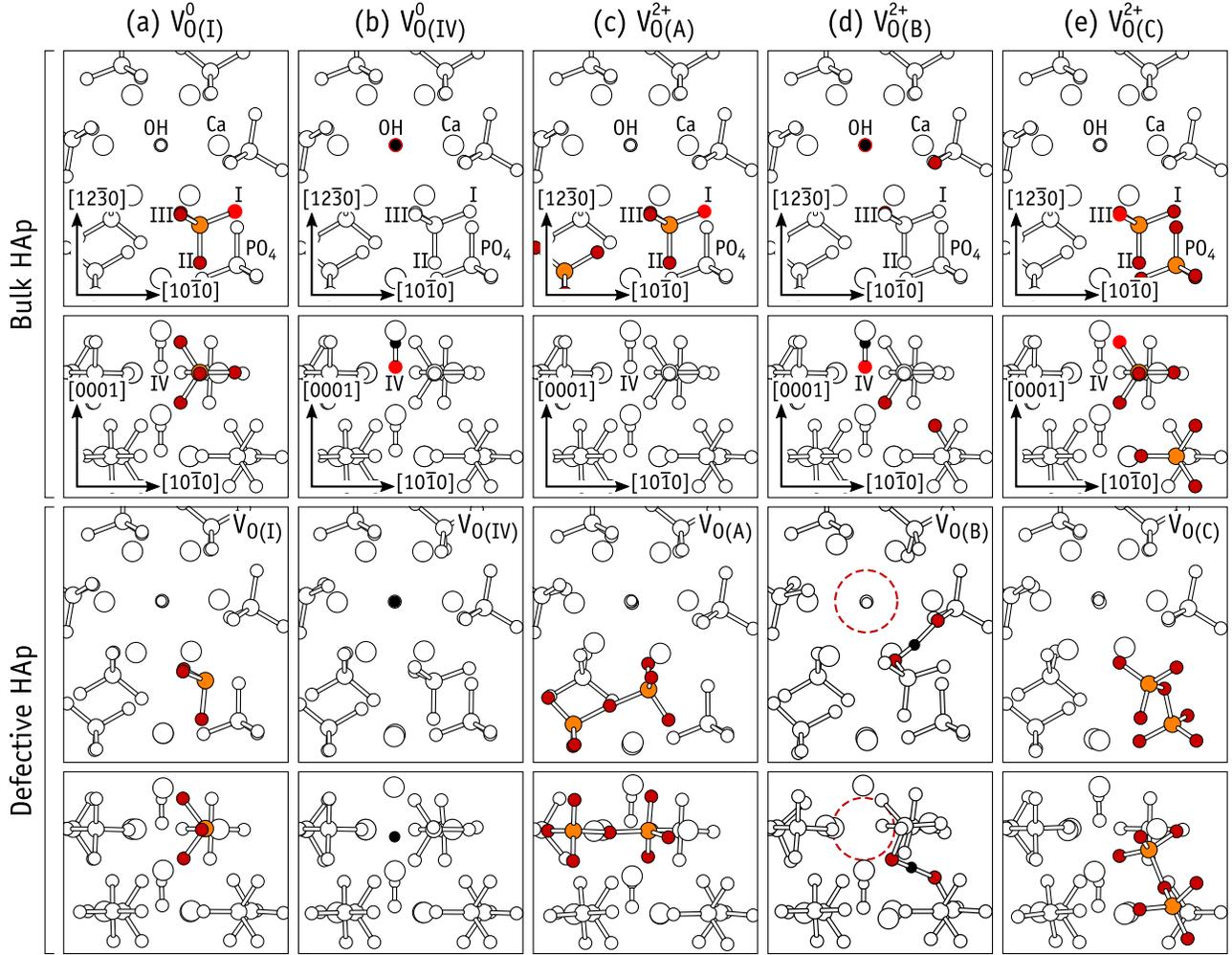}

\caption{\label{fig1}Atomistic diagrams showing ``Bulk'' (upper half) and
``Defective'' (lower half) HAp. For each row pair, upper and lower
rows display the same structures viewed along $[0001]$ and $[12\bar{3}0]$
directions of the hexagonal lattice, respectively. Formation of structures
I, IV, A, B and C of V$_{\mathrm{O}}$ defects is explained in columns
(a)-(e). Only atoms belonging to the core of the defect are colored
(P, O and H atoms are shown in orange, red and black, respectively).
Vacancies were created by removing the bright-red O-atom shown in
the ``Bulk'' figures. Upon atomic relaxation, the resulting structures
are those in the corresponding ``Defective'' figures.}
\end{figure*}

Crystalline HAp has oxygen atoms on four symmetry-inequivalent sites,
which are referred to as oxygen types I-IV. Types I, II and III are
in the phosphate units, while type IV oxygen atoms are located in
the OH$^{-}$ anions. These are denoted as O(I),...,O(IV) and are
shown in the upper half of Figure~\ref{fig1}, where portions of
bulk HAp are depicted. We note that PO$_{4}$ groups have two nearly
symmetric O(III) atoms which are superimposed in the upper view of
bulk HAp in Figure~\ref{fig1}.

Before describing the structure of the oxygen vacancies it is convenient
to leave a few words about notation. An oxygen vacancy in charge state
$q$ is denoted as V$_{\mathrm{O}(R)}^{q}$, where $R$ stands for
a particular structure label. It is important to understand that $q$
is the net charge of the whole defect (enclosing its core and possibly
many ligands) and must be distinguished from the oxidation state of
the moieties that make the core of the defect. Access to such information
would be provided with a description of the electronic state of the
defect, preferably using a compact recipe. For instance, the creation
of a neutral oxygen vacancy in the HAp lattice can be described as
a replacement of a $\mathrm{PO_{4}^{3-}}$ moiety in pristine HAp
by a $\mathrm{PO_{3}^{3-}}$ anion, both in the $-3$ oxidation state,
thus ensuring charge neutrality of the whole sample. Along these lines,
we introduce a more informative defect notation, namely $[X^{n}]_{Y}^{q}$,
which represents a defect moiety $X$ in oxidation state $n$, replacing
a crystalline site/moiety $Y$ with oxidation state $m=n-q$, thus
leading to a defect with net charge $q$. For instance, a positively
charged oxygen vacancy can either be referred to as V$_{\mathrm{O}}^{+}$
or alternatively as a $\mathrm{PO_{3}^{2-}}$ anion replacing a $\mathrm{PO_{4}^{3-}}$
moiety, \emph{i.e.}, $[\mathrm{PO_{3}^{2-}}]_{\mathrm{PO_{4}}}^{+}$.

\begin{figure}
\includegraphics[width=8.5cm]{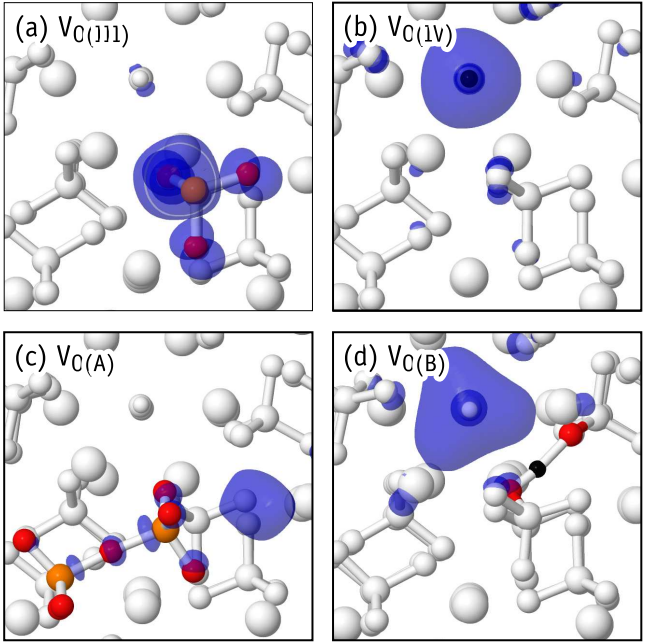}

\caption{\label{fig2}Electron density isosurfaces from the highest occupied
Kohn-Sham level of neutral V$_{\mathrm{O}}$ defects in HAp. The density
of V$_{\mathrm{O(III)}}$ in (a) is representative of V$_{\mathrm{O(I)}}$
and V$_{\mathrm{O(II)}}$ as well. Isosurfaces are drawn at constant
electron density $n=0.001$~Å$^{-3}$.}
\end{figure}

Among the many vacancy structures investigated, those shown in Figures~\ref{fig1}(a)-(e)
are the most relevant as they showed lower energy. Metastable structures
with more than $2$~eV above the ground state (for each particular
charge state) will not be discussed. Figure~\ref{fig1}(a) describes
the formation of a pyramidal PO$_{3}$ structure, where removal of
the O(I) atom in ``Bulk Hap'' highlighted using a bright red color,
leads to V$_{\mathrm{O(I)}}$ as shown in the lower part of the figure,
where a ``Defective HAp'' region is shown. Analogous structures
for V$_{\mathrm{O(II)}}$ and V$_{\mathrm{O(III)}}$ were obtained
as well. In the neutral charge state the resulting $[\mathrm{PO_{3}^{3-}}]_{\mathrm{PO_{4}}}^{0}$
structures display a fully occupied sp$^{3}$ orbital on the P atom,
resembling the phosphine molecule. This is shown in Figure~\ref{fig2}(a),
where an isosurface of the electron density corresponding to the highest
occupied state is represented in blue for the specific case of V$_{\mathrm{O(III)}}^{0}$.
Figure~\ref{fig1}(b) shows the case of a missing O(IV) atom, leaving
an isolated H atom in the OH channel. This defect can be rationalized
as the removal of neutral O from OH$^{-}$, leaving an hydride anion
to maintaining charge neutrality of the whole system. In the neutral
charge state we have a $[\mathrm{H^{-}}]_{\mathrm{OH}}^{0}$ structure,
where after structural relaxation the hydride species becomes located
close to the site of the missing O(IV) atom (compare lower diagrams
of bulk and defective HAp in Figure~\ref{fig1}(b)). The density
corresponding to the highest occupied state of V$_{\mathrm{O(IV)}}^{0}$
is represented in \ref{fig2}(b), which clearly shows the formation
of an hydride anion in the OH-channel.

In V$_{\mathrm{O(I)}}$-V$_{\mathrm{O(IV)}}$ defects, all atoms (but
the missing oxygen) remain essentially close to their original crystalline
coordinates, hence the use of subscripted O(I)-O(IV) labels to identify
their structure. However, additional structures, hereafter referred
to as \emph{extended} structures, were also found for the oxygen vacancy
in HAp. One type of such extended structures can be described as a
pair of neighboring oxygen vacancies connected by an O-interstitial,
$2\mathrm{V}_{\mathrm{O}}+\mathrm{O}$. A second type is best described
as a complex made of an OH-vacancy next to an H-interstitial, $\mathrm{V}_{\mathrm{OH}}+\mathrm{H}$.
Two defects of type $2\mathrm{V}_{\mathrm{O}}+\mathrm{O}$ are singled
out and labeled V$_{\mathrm{O(A)}}$ and V$_{\mathrm{O(C)}}$. They
are shown in Figures~\ref{fig1}(c) and \ref{fig1}(e). One $\mathrm{V}_{\mathrm{OH}}+\mathrm{H}$
defect is shown in Figure~\ref{fig1}(d) and is referred to as V$_{\mathrm{O(B)}}$.
A dashed circle is used in the figure to highlight the missing OH
unit. The highest occupied state of the extended structures is shown
in Figures~\ref{fig2}(c) and \ref{fig2}(d). They either overlap
void regions of the HAp crystal or the vacant volume of the OH-channel,
thus suggesting that they are donors with anti-bonding character or
strong resonance with conduction band states.\cite{Avakyan2018}

We found that the structure and stability of V$_{\mathrm{O}}$ defects
strongly depend on the charge state. For neutral and single positive
vacancies, only V$_{\mathrm{O(I)}}$-V$_{\mathrm{O(IV)}}$ are stable.
In these charge states, extended configurations (V$_{\mathrm{O(A)}}$-V$_{\mathrm{O(C)}}$)
relax to a V$_{\mathrm{O(I)}}$-V$_{\mathrm{O(IV)}}$ structure. Conversely,
for charge state $+2$, structures V$_{\mathrm{O(I)}}$-V$_{\mathrm{O(IV)}}$
are unstable and relax to an extended configuration.

For relaxations in charge state $+2$ that started from structures
I and III the final structures were respectively A and C. Here, the
P atom of the PO$_{3}$ unit in V$_{\mathrm{O(I)}}$ (or V$_{\mathrm{O(III)}}$)
moved across the plane defined by the three O atoms to connect to
the O atom from the nearest PO$_{4}$ moiety. Such a severe relaxation
can be explained by electron transfer from a neighboring PO$_{4}^{-3}$
anion to the empty P(sp$^{3}$) orbital of $[\mathrm{PO_{3}^{-}}]_{\mathrm{PO_{4}}}^{2+}$
in V$_{\mathrm{O(I)}}^{2+}$ or V$_{\mathrm{O(III)}}^{2+}$ (see Figure~\ref{fig2}(a)),
and the subsequent formation a new P-O bond. The result is an extended
$[\mathrm{PO_{3}^{2-}}\textrm{-}\mathrm{O}\textrm{-}\mathrm{PO_{3}^{2-}}]_{\mathrm{2(PO_{4})}}^{2+}$
structure shown in Figures~\ref{fig1}(c) and \ref{fig1}(e).

When initiating the relaxation in charge state $+2$ from structures
II and IV, the resulting configuration was in both cases V$_{\mathrm{O(B)}}^{2+}$.
In this charge state, the defect comprises an interstitial $\mathrm{H^{+}}$
next to a positively charged OH-vacancy, \emph{i.e.}, $[\mathrm{PO_{4}^{3-}}\textrm{-}\mathrm{H^{+}}\textrm{-}\mathrm{PO_{4}^{3-}}]_{\mathrm{2PO_{4}}}^{+}+\mathrm{V}_{\mathrm{OH}}^{+}$.
The proton is located on a high electron density site between two
oxygen anions. The net positive charge of the OH-vacancy follows from
depletion of two electrons from the channel-state represented by the
isosurface of Figure~\ref{fig2}(d). When starting from structure
II, Coulomb attraction and subsequent reaction between neighboring
$\mathrm{OH}^{-}$ and $[\mathrm{PO_{3}^{-}}]_{\mathrm{PO_{4}}}^{2+}$
leads to the formation of the V$_{\mathrm{O(B)}}^{2+}$ extended structure.
Alternatively, when starting from structure IV, a proton in the initial
$[\mathrm{H^{+}}]_{\mathrm{OH}}^{2+}$ configuration is strongly attracted
by O-anions in neighboring $\mathrm{PO}_{4}^{3-}$ moieties, also
ending up in V$_{\mathrm{O(B)}}^{2+}$ as depicted in Figure~\ref{fig1}(d).

Upon relaxation of neutral and positively charged structures A and
C, the final defect configurations were I and III, respectively. For
neutral and positively charged relaxations of structure B, the resulting
structure depends on the charge state. When starting from V$_{\mathrm{O(B)}}^{+}$,
the V$_{\mathrm{OH}}$ state of Figure~\ref{fig2}(d) is partially
occupied. The O-H$^{+}$-O unit dissociates spontaneously, with the
proton jumping into the OH channel to overlap with the V$_{\mathrm{OH}}$
electron and form neutral hydrogen,

\[
[\mathrm{PO_{4}^{3-}}\textrm{-}\mathrm{H^{+}}\textrm{-}\mathrm{PO_{4}^{3-}}]_{\mathrm{2PO_{4}}}^{+}+\mathrm{V}_{\mathrm{OH}}^{0}\rightarrow[\mathrm{H^{0}}]_{\mathrm{OH}}^{+}+2\mathrm{PO_{4}},
\]
ending up in the V$_{\mathrm{O(IV)}}^{+}$ state. Conversely, when
starting from V$_{\mathrm{O(B)}}^{0}$, the state shown in Figure~\ref{fig2}(d)
becomes fully occupied, so that the OH-vacancy is negatively charged.
During atomic relaxation, an $(\mathrm{OH})^{+}$ unit breaks away
from the O-H$^{+}$-O structure to fill in the nearby OH-vacant site.
The mechanism is therefore 
\[
[\mathrm{PO_{4}^{3-}}\textrm{-}\mathrm{H^{+}}\textrm{-}\mathrm{PO_{4}^{3-}}]_{\mathrm{2PO_{4}}}^{+}+\mathrm{V}_{\mathrm{OH}}^{-}\rightarrow[\mathrm{PO_{3}^{3-}}]_{\mathrm{PO_{4}}}^{0}+\mathrm{OH}+\mathrm{PO}{}_{4},
\]
now ending up in the V$_{\mathrm{O(II)}}^{0}$ state.

Relative defect energies in all charge states of interest are reported
in Table~\ref{tab1}, which includes both PBE- and B3LYP-level results.
It is interesting to note that within the same charge state, both
PBE and B3LYP data display quite a consistent picture, with relative
energies differing by less than 0.2~eV. However, as we will show
below, that is not the case when we compare electronic transition
energies obtained using the semi-local and hybrid density functionals.

In broad terms, Table~\ref{tab1} shows that neutral V$_{\mathrm{O(IV)}}^{0}$
in HAp is a distinct ground state. Pyramidal V$_{\mathrm{O(I)}}^{0}$-V$_{\mathrm{O(III)}}^{0}$
defects are at least 1~eV higher in the energy scale. Single positively
charged V$_{\mathrm{O}}^{+}$ is most stable when adopting the pyramidal
structure III. Other competing states with analogous configurations
(I and II) are metastable by about 0.3~eV. For the double plus V$_{\mathrm{O}}^{2+}$
defect, the ground state is structure C (see Figure~\ref{fig1}(e)).
The distance between the inter-planar phosphorous atoms in bulk HAp
is 4.1~Å, which is shorter than the intra-planar P$\cdots$P distance
(4.8~Å). The shorter inter-planar distance favors the formation of
V$_{\mathrm{O(C)}}^{2+}$ (in detriment of V$_{\mathrm{O(A)}}^{2+}$)
by conferring some p-character to the lone states of the bridging
O atom, and making V$_{\mathrm{O(A)}}^{2+}$ metastable by at least
1.3~eV. This is evident in Figures~\ref{fig1}(c) and \ref{fig1}(e),
that show P-O-P angles of 175° and 125°, respectively. Finally, we
found a competing double positive state, namely structure B, only
0.2~eV above the ground state. As described in Figure~\ref{fig1}(d),
structure B comprises an OH vacancy next to H interstitial. Interestingly,
the energy of infinitely separated V$_{\mathrm{OH}}^{+}$ and H$^{+}$
defects (obtained from independent supercell calculations) is 0.2~eV
lower than that of V$_{\mathrm{O(B)}}^{++}$, being essentially degenerate
to V$_{\mathrm{O(C)}}^{++}$. In these calculations, the isolated
proton was found to be more stable when located between two O(III)
atoms, forming a O-H-O structure with slightly asymmetric OH distances
(analogously to Figure~\ref{fig1}(d)).

Although further work is necessary, we may speculate that once a V$_{\mathrm{O(B)}}^{++}$
defect is formed, the separation of V$_{\mathrm{OH}}^{+}$ and H$^{+}$is
most likely to take place via consecutive jumps of neighboring OH$^{-}$
anions into the positively charged OH-vacancy.

\begin{table}
\caption{\label{tab1}Relative energies of V$_{\mathrm{O}}$ defects in HAp
with respect to the lowest energy structure in a particular charge
state (highlighted in bold). Results from both PBE and B3LYP calculations
are reported. For unstable structures, the energy is replaced by the
label of the respective structure obtained after relaxation. All values
are in eV.}

\begin{tabular}{ccccc}
\hline 
Structure & Functional & V$_{\mathrm{O}}^{0}$ & V$_{\mathrm{O}}^{+}$ & V$_{\mathrm{O}}^{++}$\tabularnewline
\hline 
\multirow{2}{*}{I} & PBE & 1.01 & 0.33 & \multirow{2}{*}{A}\tabularnewline
 & B3LYP & 1.17 & 0.35 & \tabularnewline
\multirow{2}{*}{II} & PBE & 0.96 & 0.23 & \multirow{2}{*}{B}\tabularnewline
 & B3LYP & 1.14 & 0.34 & \tabularnewline
\multirow{2}{*}{III} & PBE & 0.88 & \textbf{0.00} & \multirow{2}{*}{C}\tabularnewline
 & B3LYP & 1.03 & \textbf{0.00} & \tabularnewline
\multirow{2}{*}{IV} & PBE & \textbf{0.00} & 0.67 & \multirow{2}{*}{B}\tabularnewline
 & B3LYP & \textbf{0.00} & 0.54 & \tabularnewline
\multirow{2}{*}{A} & PBE & \multirow{2}{*}{I} & \multirow{2}{*}{I} & 1.21\tabularnewline
 & B3LYP &  &  & 1.35\tabularnewline
\multirow{2}{*}{B} & PBE & \multirow{2}{*}{II} & \multirow{2}{*}{IV} & 0.17\tabularnewline
 & B3LYP &  &  & 0.19\tabularnewline
\multirow{2}{*}{C} & PBE & \multirow{2}{*}{III} & \multirow{2}{*}{III} & \textbf{0.00}\tabularnewline
 & B3LYP &  &  & \textbf{0.00}\tabularnewline
\hline 
\end{tabular}
\end{table}

\begin{figure}
\includegraphics[width=8.5cm]{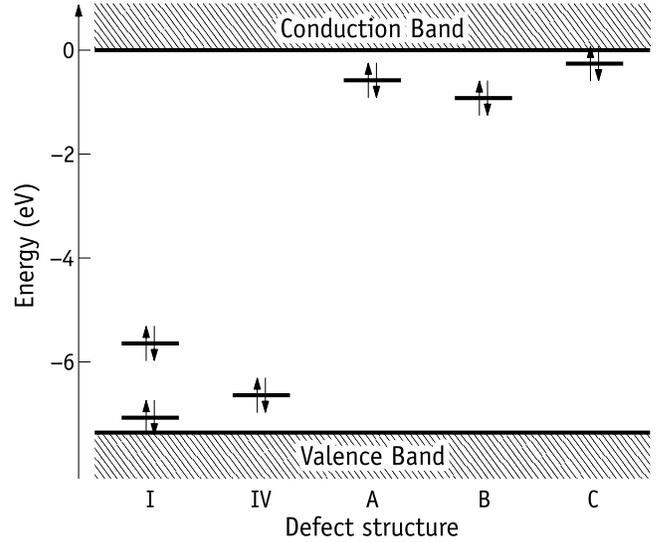}

\caption{\label{fig3}Kohn-Sham energy levels of neutral V$_{\mathrm{O}}$
defects in a HAp supercell at the $\mathbf{k}=\Gamma$ point. Defect
structure I is also representative of structures II and III (see text).
The latter have gap states that deviate from those of V$_{\mathrm{O(I)}}$
by less than 0.2~eV.}
\end{figure}

Defect structures described above were only investigated in neutral,
positive and double positive charge states. Inspection of the Kohn-Sham
eigenvalues at $\mathbf{k}=\Gamma$ confirmed that V$_{\mathrm{O}}$
defects are all donors. The diagram in Figure~\ref{fig3} shows the
B3LYP Kohn-Sham energies of neutral vacancies in the energy range
of the one-electron band gap $\epsilon_{\mathrm{cb}}-\epsilon_{\mathrm{vb}}=7.34$~eV.
Here $\epsilon_{\mathrm{cb}}$ and $\epsilon_{\mathrm{vb}}$ are energies
of the bottom and top of the conduction band and valence band, respectively.
These are represented by solid horizontal lines spanning the whole
diagram width. Short horizontal lines represent the defect-related
states, as obtained from the Kohn-Sham eigenvalues. Level occupation
is indicated by upward and downward arrows. The value of $\epsilon_{\mathrm{cb}}$
was chosen to be at the origin of the energy scale. We note that since
structures A-C are only stable in the double plus charge state, their
one electron energies were obtained from neutral single-point calculations
using V$_{\mathrm{O(A)}}^{2+}$-V$_{\mathrm{O(C)}}^{2+}$ atomic structures.

The electronic structure of V$_{\mathrm{O(II)}}$ and V$_{\mathrm{O(III)}}$
were essentially identical to that of V$_{\mathrm{O(I)}}$. Considering
that all diagrams refer to neutral states, it becomes evident that
V$_{\mathrm{O}}$ defects show no empty gap states, and therefore
cannot trap electrons. Hence, V$_{\mathrm{O}}$ defects in HAp are
not stable in the negative charge state.

Although the Kohn-Sham band structure lacks physical significance,
we can always make use of Koopmans' theorem to connect the highest
occupied state to the photo-ionization energy of the system under
scrutiny.\cite{Koopmans1934,Perdew1982} With this in mind, it is
evident that the simpler V$_{\mathrm{O(I)}}$-V$_{\mathrm{O(IV)}}$
structures are deep donors with photo-ionization energies close to
6~eV, while semi-vacancy configurations are shallower donors, where
promotion of electrons to the conduction band requires a low ionization
energy. This makes V$_{\mathrm{O(A)}}^{2+}$, V$_{\mathrm{O(B)}}^{2+}$
and V$_{\mathrm{O(C)}}^{2+}$ rather stable species.

The energy of the highest occupied states in Figure~\ref{fig3} account
for \emph{vertical} transitions which are in principle accessible
by optical excitation. However, transition levels (or electronic levels)
of defects are equilibrium properties and must be derived from the
formation energy of both ground states involved in the transitions.
This is analogous to the zero-phonon (ZP) line energy in photoluminescence/absorption
experiments. Hence, the energy of a transition level of a defect will
differ from the vertical (optical) transition by a Franck-Condon relaxation
energy. These issues will be discussed below.

Equation~\ref{eq:formation} allows us to construct a \emph{phase-diagram}
for the oxygen vacancy in HAp as a function of the chemical potentials
and Fermi energy. This is particularly useful for defects which adopt
different structures in several charge states, allowing us to immediately
identify the most stable states for a particular position of the Fermi
energy. The value of $E_{\mathrm{F}}$ for which two solutions $E_{\mathrm{f}}(q)$
and $E_{\mathrm{f}}(q+1)$ from Eq.~\ref{eq:formation} become identical,
defines a threshold above which V$_{\mathrm{O}}^{q}$ is more stable
than V$_{\mathrm{O}}^{q+1}$. This is the defect transition level
and can be calculated with respect to the valence band top as,

\begin{equation}
E(q/q+1)-E_{\mathrm{v}}=[E(q,R^{q})-E(q+1,R^{q+1})]-I_{\mathrm{bulk}},\label{eq:level}
\end{equation}
where we distinguish any possible different structures $R^{q}$ and
$R^{q+1}$ for charge states $q$ and $q+1$, respectively, \emph{E}
is the total energy (\emph{cf.} Eq.~\ref{eq:formation}), which incorporates
a periodic charge correction for $q\neq0$, and $I_{\mathrm{bulk}}=E_{\mathrm{bulk}}(q=+1)-E_{\mathrm{bulk}}(q=0)$
is the ionization energy of a bulk supercell. Note that alternatively,
one could have replaced $I_{\mathrm{bulk}}$ in Eq.~\ref{eq:level}
by the energy of the highest occupied Kohn-Sham level from a bulk
calculation. We have chosen to avoid the use of Kohn-Sham energies,
keeping the results solely based on total energies. The Kohn-Sham
band gap from B3LYP-level calculations ($E_{\mathrm{g}}^{\mathrm{KS}}=7.34$~eV)
is about 0.49~eV wider than the quasi-particle gap (calculated as
$E_{\mathrm{g}}^{\mathrm{QP}}=E_{\mathrm{bulk}}(+1)+E_{\mathrm{bulk}}(-1)-2E_{\mathrm{bulk}}(0)=6.85$~eV).

\begin{figure}
\includegraphics[width=8.5cm]{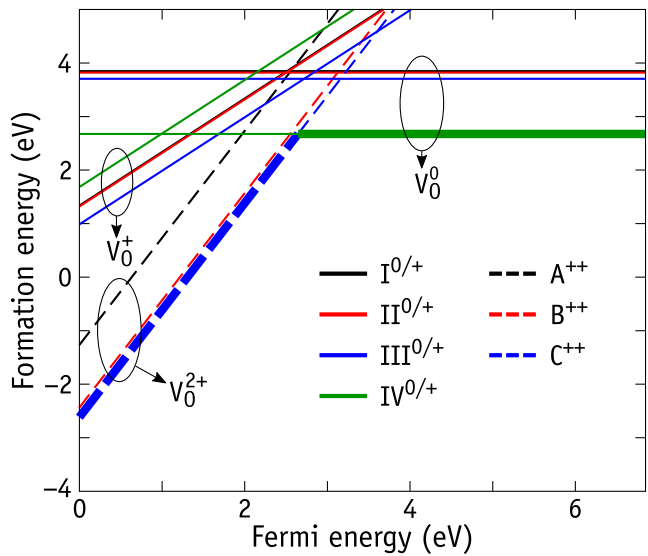}

\caption{\label{fig4}Formation energy diagram of the V$_{\mathrm{O}}$ defect
in HAp as a function of the Fermi energy. Solid lines represent formation
energies of neutral and single positive defects with structures I-IV,
while dashed lines represent double positive defects with structures
A-C.}
\end{figure}

Figure~\ref{fig4} plots the formation energy for several V$_{\mathrm{O}}$
defects in different charge states according to Eq.~\ref{eq:formation}.
The horizontal axis represents the Fermi energy with $E_{\mathrm{F}}=0$~eV
representing the valence band top and $E_{\mathrm{F}}=6.85$~eV the
conduction band bottom. Horizontal, single positive and double positive
sloped lines represent formation energies of V$_{\mathrm{O}}^{0}$,
V$_{\mathrm{O}}^{+}$ and V$_{\mathrm{O}}^{2+}$, respectively. Solid
lines stand for structures I-IV (in charge states $q=0$ and $q=+1$),
while dashed lines denote extended structures A, B and C (see legend).
Thick lines highlight the lowest energy states, namely V$_{\mathrm{O(IV)}}^{0}$
and V$_{\mathrm{O(III)}}^{2+}$, for any position of $E_{\mathrm{F}}$.

In line with the discussion above, among the formation energies of
neutral defects (horizontal lines), that of V$_{\mathrm{O(IV)}}^{0}$
is lowest, with other structures being metastable by about 1~eV.
For the positive charge state, pyramidal $[\mathrm{PO}_{3}^{2-}]_{\mathrm{PO_{4}}}^{+}$
defects are more stable. In this case, structure III has lower energy,
but structures I and II are metastable by only 0.35~eV. For double
positively charged structures, V$_{\mathrm{O(B)}}^{2+}$ and V$_{\mathrm{O(C)}}^{2+}$
are the most stable, with a slight preference for V$_{\mathrm{O(C)}}^{2+}$.

It is important to realize that under no circumstance the positive
charge states are thermodynamically stable (irrespectively of the
Fermi energy location). Formation energies of V$_{\mathrm{O}}^{+}$
defects are always higher than those of neutral and double positive
defects. This property is usually referred to as negative-$U$,\cite{Anderson1975,Watkins1984}
alluding to the negative correlation energy between three consecutive
charge states. This is best illustrated if compared to the case of
isolated atoms --- here the $n$-th ionization energy ($I_{n}$)
is smaller than the $(n+1)$-th ionization energy ($I_{n+1}$), where
the correlation between the first ionized state and remaining electrons
is $U=I_{n+1}-I_{n}>0$. This is mostly due a decrease of electron-electron
repulsion upon ionization (and a corresponding increase of electron-nuclei
Coulomb attraction due to lower electronic screening). However, for
some defects in crystals, as well as for some molecules, the first
ionization leads to a strong atomic relaxation/transformation (not
possible in single atoms), allowing a trade between the energy from
bond formation/breaking, and the Coulomb energy which binds the ionized
state. For sufficiently large relaxations, the second ionization energy
may actually become smaller than the first, and the defect may effectively
show a negative-$U$.

A consequence of a negative-$U$ ordering of donor levels (as depicted
in Figure~\ref{fig4}), is an entangled promotion of two electrons
to the conduction band when a neutral V$_{\mathrm{O}}^{0}$ defect
is excited by means of optical or thermal excitations. Analogously,
when the defect is fully ionized and there are free electrons available
in the conduction band (upon exposure to ultra-violet light), the
capture of the first electron (into V$_{\mathrm{O}}^{+}$) will make
the defect even more \emph{eager} for a second capture (into V$_{\mathrm{O}}^{0}$).

Another important feature is the fact that under thermal equilibrium,
single positive defects are not stable. Considering the lowest energy
structures for each charge state, namely V$_{\mathrm{O(IV)}}^{0}$,
V$_{\mathrm{O(III)}}^{+}$ and V$_{\mathrm{O(III)}}^{2+}$, we always
have $E_{\mathrm{f}}(2+)+E_{\mathrm{f}}(0)<2E_{\mathrm{f}}(+)$, and
only V$_{\mathrm{O(IV)}}^{0}$ and V$_{\mathrm{O(C)}}^{2+}$ (and/or
possibly V$_{\mathrm{O(B)}}^{2+}$) may occur. These states are represented
by thick lines in Figure~\ref{fig4}. It is clear that they define
a $(0/2+)$ transition level at the crossing point. The location of
the level is calculated at $E(0/2+)-E_{\mathrm{v}}=2.65$~eV. Hence,
for p-type HAp, where the Fermi energy is at the lower-half of the
gap, oxygen vacancies are likely to adopt double positive extended
structures V$_{\mathrm{O(B)}}^{2+}$ or V$_{\mathrm{O(C)}}^{2+}$.
Conversely, in n-type (and intrinsic) HAp the neutral state is more
stable and the defect will be found as an isolated H$^{-}$ hydride
in the OH-channel (\emph{i.e.}, V$_{\mathrm{O(IV)}}^{0}$).

For the sake of reference, we leave the reader with Table~\ref{tab2},
where the calculated transition levels, $\Delta E(q/q+1)=E(q/q+1)-E_{\mathrm{v}}$
involving the relevant structures $R^{q}$ and $R^{q+1}$, can be
found. Results based on both PBE and B3LYP functionals are reported.
Unlike the energy differences in the same charge state (\emph{c.f.}
Table~\ref{tab1}), it is evident that PBE results have error bars
of at least 1~eV, which is in line with our findings reported in
Ref.~\citenum{Avakyan2018}.

\begin{table*}
\caption{\label{tab2}Calculated electronic donor levels of V$_{\mathrm{O}}$
in HAp with respect to the valence band top, $\Delta E(q/q+1)=E(q/q+1)-E_{\mathrm{v}}$.
Both semi-local DFT (PBE) and hybrid-DFT (B3LYP) results are reported.
Also indicated are defect structures, $R^{q}$, corresponding to each
charge state in the transition. All values are in eV.}

\begin{tabular}{ccccc}
\hline 
Functional & $\Delta E(0/\!+)$ & $R^{0}/R^{+}$ & $\Delta E(+/2+)$ & $R^{+}/R^{2+}$\tabularnewline
\hline 
PBE & 1.53 & \multirow{2}{*}{I/I} & 1.71 & \multirow{2}{*}{I/A}\tabularnewline
B3LYP & 2.51 &  & 2.60 & \tabularnewline
PBE & 1.58 & \multirow{2}{*}{II/II} & 2.64 & \multirow{2}{*}{II/B}\tabularnewline
B3LYP & 2.50 &  & 3.74 & \tabularnewline
PBE & 1.72 & \multirow{2}{*}{III/III} & 2.59 & \multirow{2}{*}{III/C}\tabularnewline
B3LYP & 2.72 &  & 3.60 & \tabularnewline
PBE & 0.17 & \multirow{2}{*}{IV/IV} & 3.08 & \multirow{2}{*}{IV/B}\tabularnewline
B3LYP & 0.99 &  & 3.94 & \tabularnewline
\hline 
\end{tabular}
\end{table*}

\section{Discussion\label{sec:discussion}}

Our results suggest that under UV illumination, the most likely excitation
channels for the oxygen vacancy are (i) electron emission from V$_{\mathrm{O(IV)}}^{0}$
to the conduction band, (ii) hole-emission from V$_{\mathrm{O(B)}}^{2+}$
or V$_{\mathrm{O(C)}}^{2+}$ to the valence band (which is equivalent
to the promotion of an electron from the valence band to the empty
donor state of a double positive vacancy), or even (iii) electron
emission from metastable V$_{\mathrm{O(I)}}^{0}$-V$_{\mathrm{O(III)}}^{0}$
defects.

For the first case, the energy needed to perform a vertical transition
from the ground state structure,

\[
\mathrm{V}_{\mathrm{O(IV)}}^{0}\xrightarrow{\textrm{vertical}}\mathrm{V}_{\mathrm{O(IV)}}^{+}+e^{-}
\]
is estimated to be higher than $E_{\mathrm{ph}}\sim6.5$~eV (see
Kohn-Sham levels for structure IV in Figure~\ref{fig3}), where $e^{-}$
represents an electron at the conduction band bottom. This energy
is far too large to be connected with the optical excitation threshold
observed during photocatalysis in the range 3.4-4.0~eV.\cite{Bystrov2016}

As previously referred, the calculated electronic levels of Table~\ref{tab2}
provide us access to zero-phonon (adiabatic) transition energies.
For a neutral vacancy, adiabatic electron emission (ionization) energies
are given by $E_{\mathrm{ad}}=E_{\mathrm{g}}-\Delta E(0/+)$. Hence,
the lowest energy transitions in absorption follow

\[
\mathrm{V}_{\mathrm{O(IV)}}^{0}\xrightarrow{\textrm{adiabatic}}\mathrm{V}_{\mathrm{O(I,II,III)}}^{+}+e^{-},
\]
corresponding to an absorption energy $E_{\mathrm{ad}}=5.51$~eV,
5.49~eV, 5.16~eV for transitions where the final structures are
V$_{\mathrm{O(I)}}^{+}$, V$_{\mathrm{O(II)}}^{+}$, and V$_{\mathrm{O(III)}}^{+}$,
respectively. In the above, $\Delta E(0/+)=1.34$~eV, 1.36~eV and
1.69 are donor transition energies considering $R^{0}=\textrm{IV}$
and $R^{+}=\text{I}$, II and III, respectively, while $E_{\mathrm{g}}=6.85$~eV
is the calculated quasi-particle gap width. Although the adiabatic
transition energies are closer to the experimental data, kinetic effects
are expected limit the exchange of oxygen between neighboring PO$_{4}$
and OH moieties,

\[
[\mathrm{H^{-}}]_{\mathrm{OH}}^{0}+\mathrm{PO}_{4}\rightarrow\mathrm{OH}+[\mathrm{PO_{3}^{2-}}]_{\mathrm{PO_{4}}}^{+}+e^{-},
\]
where the first reactant and second product represent V$_{\mathrm{O(IV)}}^{0}$
and V$_{\mathrm{O(I,II,III)}}^{+}$ defects, respectively.

We also note that being a negative-$U$ defect, under persistent excitation
a second ionization of V$_{\mathrm{O(I,II,III)}}$ would immediately
follow the first, but now with considerably lower energy, \emph{i.e.}
in the range 3.3-4.3~eV, leading to the formation of V$_{\mathrm{O(A,B,C)}}^{2+}$
defects.

Mechanism (ii), namely a hole-emission from V$_{\mathrm{O(C)}}^{2+}$
or from V$_{\mathrm{O(B)}}^{2+}$, would be likely in p-type HAp,
where the Fermi energy is in the lower half of the band gap and the
ground state is double positive (see Figure~\ref{fig4}). Vertical
transitions for both B and C structures follow

\[
\mathrm{V}_{\mathrm{O}}^{2+}\xrightarrow{\textrm{vertical}}\mathrm{V}_{\mathrm{O}}^{+}+h^{+},
\]
where $h^{+}$ stands for a hole in the valence band, and the energy
as estimated from the B3LYP one-electron energies is close to 7~eV.
Such a large value results from the shallow donor nature of the extended
defects, which place the donor transition very high in the gap (see
Figure~\ref{fig3}).

If we instead consider an adiabatic hole emission from V$_{\mathrm{O(C)}}^{2+}$
to the lowest energy V$_{\mathrm{O}}^{+}$ structure, \emph{i.e.}
V$_{\mathrm{O(III)}}^{+}$, we find 3.6~eV, in rather good agreement
with the optical experiments.\cite{Bystrov2016} This mechanism involves
a transition between ground states (in their respective charge state),
consisting on breaking one of the P-O bonds in the O$_{3}$P-O-PO$_{3}$
extended structure of V$_{\mathrm{O(C)}}^{2+}$.

The adiabatic transition from V$_{\mathrm{O(B)}}^{2+}$ to V$_{\mathrm{O(II)}}^{+}$
involves an absorption of 3.7~eV, but in this case the mechanism
implies the a displacement of OH$^{-}$ from the V$_{\mathrm{O(B)}}^{2+}$
structure (see Figures \ref{fig1}(e)) to the OH-channel of the HAp
lattice. Alternatively, a transition from V$_{\mathrm{O(B)}}^{2+}$
to V$_{\mathrm{O(IV)}}^{+}$ simply involves a jump of H from the
O-H-O unit in V$_{\mathrm{O(B)}}^{2+}$ to a close OH site in V$_{\mathrm{O(IV)}}^{+}$,

\[
[\mathrm{PO_{4}^{3-}}\textrm{-}\mathrm{H^{+}}\textrm{-}\mathrm{PO_{4}^{3-}}]_{\mathrm{2PO_{4}}}^{+}+\mathrm{V}_{\mathrm{OH}}^{+}\rightarrow[\mathrm{H}^{0}]_{\mathrm{OH}}^{+}+2\mathrm{PO_{4}}+h^{+},
\]
which we find to proceed spontaneously (without an impeding barrier)
upon hole emission. In the transition above, the reactants and first
product represent V$_{\mathrm{O(B)}}^{2+}$ and V$_{\mathrm{O(IV)}}^{+}$
defects, respectively. The energy needed for the transformation is
3.9~eV, also accounting well for the observed UV absorption energy.

Finally, considering (iii) the photo-ionization of metastable neutral
V$_{\mathrm{O(I)}}$-V$_{\mathrm{O(III)}}$ defects (by promotion
of electrons into the conduction band), which could be present in
the material, we find again large vertical ionization energies (close
to 6~eV), whereas adiabatic energies are in the range 4.1-4.3~eV,
also close to the experimental threshold energy for the observation
of photocatalysis.

All the above transitions induce quite large changes to the local
P-O bonding. That would explain the XPS and FTIR experiments of Refs.~\citenum{Nishikawa2007},
\citenum{Reddy2007} and \citenum{Shariffuddin2013}, that show a
clear transformation of the P-O bonds upon UV illumination of HAp.
However, based on our results it is not possible to provide a final
assignment to that effect. Decisive help would be provided by combining
local vibrational mode calculations with vibrational FTIR spectroscopy
on samples with good quality. Despite the uncertainty, our results
suggest that the excitation channel (ii), which involves a photo-induced
bond-breaking of extended structures, and shows a transition energy
in excellent agreement with the observations, appears to be the strongest
candidate to explain the onset absorption of HAp in the range 3.4-4.0~eV.

\section{Conclusions\label{sec:conclusions}}

We presented a study of the structural and electronic properties of
oxygen vacancies in hydroxyapatite by means of hybrid density functional
theory within the plane-wave formalism. The vacancies were investigated
in large supercells, from which formation energies and electronic
transition energies were calculated.

We found that the vacancies essentially occur in two distinct forms,
either as a simple vacant oxygen site (referred to as structures I-IV),
or as an oxygen atom replacing two neighboring oxygen vacancies (extended
structures A-C). The former type of vacancies are deep donors, while
the latter are shallow donors with rather low ionization energies.
No acceptor states (stable negatively charged defects) were found.

Vacancy structures I-IV are more stable in the neutral charge state,
while extended structures A-C are preferred in the double plus charge
state. This means that the oxygen vacancy adopts rather different
configurations on samples where the Fermi energy is in the upper or
the lower half of the band gap. The large relaxation energies upon
ionization makes the oxygen vacancy a negative-$U$ defect, where
the single plus charge state is metastable, being only attainable
under excitation (for instance by UV illumination).

From inspection of the one-electron Kohn-Sham levels, combined with
the transition levels obtained from total energies, we find that electron
promotion from the valence band top to the donor state of the positively
charged structures, involves a zero-phonon absorption of 3.6-3.9~eV.
This transition leads to a spontaneous breaking of either P-O-P or
O-H-O bridge-bonds, and most likely explains the 3.4-4.0~eV absorption
onset for the observation of photocatalysis under persistent UV illumination.
\begin{acknowledgement}
This work was supported by the \emph{Fundação para a Ciência e a Tecnologia}
(FCT) through project UID/CTM/50025/2013 and by the Russian Foundation
for Basic Research (RFBR) through Grant No. 19-01-00519 A.
\end{acknowledgement}
\bibliography{refs}

\pagebreak{}

\begin{figure}
\includegraphics[height=4.4cm]{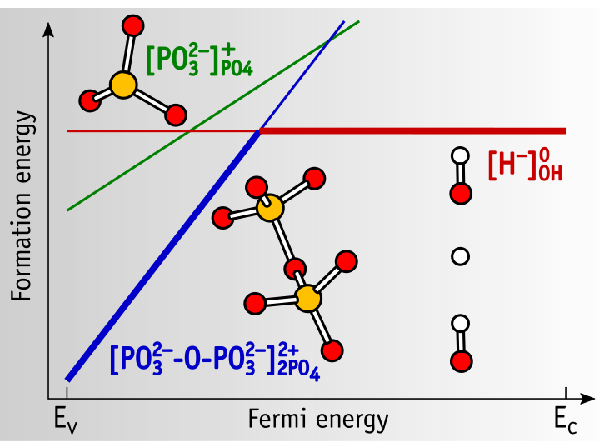}

\emph{Table of Contents Graphic}
\end{figure}

\end{document}